\documentclass[pra,amssymb,amsmath,aps,showpacs,showkeys,twocolumn]{revtex4}
\usepackage{graphicx}

\newcommand{\bc}{\begin{center}}
\newcommand{\ec}{\end{center}}

\usepackage{latexsym}
\usepackage{color}

\newcommand{\be}{\begin{equation}}
\newcommand{\ee}{\end{equation}}

\newcommand{\bex}{\begin{eqnarray}}
\newcommand{\eex}{\end{eqnarray}}
\newcommand{\bmin}{\begin{center}\begin{minipage}{460pt}}
\newcommand{\emin}{\end{minipage}\end{center}}

\begin{document}

\small

\title[The Quantum Measurement Problem and Physical 
reality]{The Quantum Measurement Problem and Physical
reality: \\A Computation Theoretic Perspective}
\thanks{
\copyright 2006 American Institute of Physics. 
The article appeared in
{\em Quantum Computing: Back Action 2006}, IIT Kanpur, India,
March 2006, Ed. Dr. D. Goswami,
AIP Conference Proceedings {\bf 864}, pp. 178--193 (2006).}
\author{R. Srikanth}
\email{srik@rri.res.in}
\affiliation{Poornaprajna Institute of Scientific Research,
Devanahalli, Bangalore 562 110, India,}
\affiliation{Raman Research Institute, Sadashiva Nagar, Bangalore- 560 080}
\thanks{Affiliation updated.}

\begin{abstract}
Is the universe computable? If yes, is it computationally a polynomial
place?   In   standard  quantum  mechanics,   which  permits  infinite
parallelism  and the  infinitely  precise specification  of states,  a
negative  answer to both  questions is  not ruled  out.  On  the other
hand, empirical evidence suggests  that {\bf NP}-complete problems are
intractable  in the physical  world. Likewise,  computational problems
known to be algorithmically uncomputable  do not seem to be computable
by  any physical  means.  We  suggest that  this  close correspondence
between the  efficiency and  power of abstract  algorithms on  the one
hand, and physical computers on the other, finds a natural explanation
if the universe  is assumed to be algorithmic;  that is, that physical
reality is the product of discrete sub-physical information processing
equivalent  to the actions  of a  probabilistic Turing  machine.  This
assumption  can  be reconciled  with  the  observed exponentiality  of
quantum systems at microscopic  scales, and the consequent possibility
of  implementing  Shor's quantum  polynomial  time  algorithm at  that
scale, provided the degree of superposition is intrinsically, finitely
upper-bounded.  If this bound is associated with the quantum-classical
divide  (the Heisenberg  cut),  a natural  resolution  to the  quantum
measurement   problem  arises.    From  this   viewpoint,  macroscopic
classicality is  an evidence  that the universe  is in {\bf  BPP}, and
both questions  raised above receive affirmative  answers.  A recently
proposed computational model of quantum measurement, which relates the
Heisenberg  cut  to the  discreteness  of  Hilbert  space, is  briefly
discussed.  A connection to quantum gravity is noted.  Our results are
compatible   with  the   philosophy  that   mathematical   truths  are
independent of the laws of physics.
\end{abstract}

\maketitle
\received{}
\revised{}
\accepted{}

\section{Introduction}
The   advance   of  quantum   information   and  quantum   computation
\cite{nc00,gruska} as a serious field  of research has brought forth a
new language  for looking at  problems in physics  and a novel  way of
characterizing   physical   theories   \cite{halv03,har01}.   Further,
information   processing  may   play   a  fundamental   role  in   our
understanding    of    physical    laws    \cite{llo02,bek03,Shiv,ng}.
Conversely,  the discovery of  Shor's algorithm  \cite{shor} suggested
that knowledge of  the physical world is relevant  to study the limits
of efficient computation.   These developments highlight the interplay
between  insights  into   information  and  computation  developed  by
computer scientists on the one hand, and well-tested physical theories
documented  by  physicists  on  the  other.  Although  the  idea  that
information theory  or computer science  can provide powerful  ways to
describe  and  explore  the  consequences  of  physical  laws  may  be
acknowledged as  intuitively appealing, belief in its  usefulness as a
guide to new physics is unlikely to be widespread among physicists. We
hope  this work  can help  elucidate  the connection  between the  two
fields.  Indeed, several independent  approaches inspired by a similar
philosophy have been undertaken \cite{fli,zei,ulf,fred}.

The eminent  physicist J. Wheeler has suggested  that physical reality
itself may  be a manifestation  of information processing  through his
idea expressed as  the aphorism ``it from bit"  \cite{hweo}.  To quote
Wheeler:
\begin{quote}
``It from  bit" symbolizes  the idea that  every item of  the physical
world has at  the bottom-- at the very bottom,  in most instances-- an
immaterial source  and explanation; that which we  call reality arises
from  the   posing  of  yes-no  questions,  and   the  registering  of
equipment-invoked  responses;  in  short,  that  things  physical  are
information theoretic in origin.
\end{quote}
In this  work, we suggest that  the known power  of physical computers
and  the quantum  measurement problem  together  offer a  clue to  and
suggest a particular realization of the above idea.

We  believe that  this approach  highlights how  computation theoretic
considerations  can  shed light  on  fundamental  questions about  why
physical laws are the way they are, and also help constrain or uncover
new  physics.  For  example, it  is  an interesting  question how  the
theoretical  model of  computation compares  with  physical computers.
For concreteness,  one can ask whether {\bf  NP}-complete problems can
be  solved in  polynomial time  using  the resources  of the  physical
universe.   What can  we say  about physics  depending on  whether the
answer  is  affirmative  or  negative?  These are  some  questions  we
consider here.

Doubtless, various profound approaches  to such questions exist, among
them string  theory, different  theories of quantum  gravity, particle
physics,  etc.   Various  recondite  clues,  such as  the  black  hole
information paradox,  dark matter, dark  energy, gamma ray  bursts are
potential  harbingers of  new physics.   Perhaps an  advantage  of the
(quantum) information and computation  theoretic approach such as this
is  that it  may  be  possible to  test  predictions using  relatively
accessible optical and quantum optical experimental techniques.

The article is  divided as follows. In the next  section, we note that
{\bf NP}-complete problems are found to be intractable in the physical
world.  This idea  can be brought into perspective  by showing that if
quantum mechanics (QM)  were not linear, or not  unitary, or not local
(ie., conforming  to the no-signalling theorem), or  not conforming to
the  Born  $|\psi|^2$  rule  \cite{nc00},  more  efficient  models  of
computation    would   be    possible   than    believed    to   exist
\cite{aar04,aar05}.    In  the  subsequent   section,  we   note  that
algorithmically uncomputable problems are believed to be unsolvable in
the physical world.  The question as to why Nature seems to be exactly
as  efficient and  powerful as  theoretical models  of  computation is
considered thereafter.   It is pointed out that  one possible solution
is  that  Nature  is  algorithmic:  that is,  physical  reality  is  a
manifestation  of discrete  sub-physical computations  and information
processing.    We   further  indicate   why   this  somewhat   unusual
interpretation  receives clarification  and support  from  the quantum
measurement   problem.    The    further   Section   recapitulates   a
computational  model  of  quantum  measurement,  compatible  with  the
conclusion  of  the  preceding  section.   Possible  implications  for
quantum  gravity are  then  discussed before  concluding  in the  last
Section.

\section{Computational complexity considerations\label{sec:np}}
In a remarkable work that founded computer science, in the course
of studying the problem of what it means to be computable, Turing formalized 
the intuitive notion of an {\em effective procedure} or
{\em algorithm} for a computational task by means of a class of abstract
devices that have come to be called
Turing machine  (TM). A TM is a symbol-manipulating device, equipped
with a movable tape of finite but unbounded length, divided into cells.
Each cell is filled with an element from a finite alphabet which includes
a blank symbol. A TM is equipped with a  
read-write head  that possesses a property known as {\em state}, a
finite set of instructions
for how the head should modify the active cell, move the tape and alter
its own state \cite{tur}.

It is of interest to study the existence
of {\em  efficient} algorithms to perform certain  tasks, the relevant
resources being (memory) space and time (i.e., number of computational
steps).  The complexity class ${\bf P}$ denotes
the set of all  problems solvable on a  TM in polynomial
time, that is,  in steps that are polynomial as  a function of problem
size. Problems   in  {\bf   P} are  often considered as the class 
of computational problems which are "efficiently solvable" or "tractable".
With the  advent of  probabilistic TMs  \cite{nc00} and
then  quantum  TMs \cite{deutsch,umesh},  
there are potentially larger classes, such as
{\bf  BPP}   (for  `bounded-error,
probabilistic,  polynomial  time) and  {\bf  BQP} (for  `bounded-error,
quantum, polynomial time), that are considered tractable.  
A probabilistic TM can be considered
as a (deterministic) TM with access to genuine randomness. {\bf BPP} denotes
the set of decision problems solvable  on a probabilistic TM in polynomial time
with   error  probability of at most $1/3$ on all instances. The choice of
$1/3$, which is arbitrary, can be replaced with any constant between
0 and $1/2$. Analogously, {\bf BQP} denotes the class of decision problems
solvable by a  quantum computer in polynomial time, with error
probability of at most $1/3$ \cite{umesh}.  Obviously, ${\bf  P}
\subseteq {\bf BPP} \subseteq {\bf BQP}$.

Roughly speaking, the set of  decision problems whose solutions can be
easily verified is  called {\bf NP}.  More precisely,  {\bf NP} is the
class of decision problems such that (a) to every positive instance of
answer,  there   exists  a  polynomial-size  proof   verifiable  by  a
deterministic   polynomial-time   algorithm   (i.e.,  an   efficiently
computable witness of membership);  and (b) to every negative instance
of answer, the algorithm must declare invalid any purported proof that
the answer is ``yes" \cite{cpa}.
An example of a problem in {\bf NP} is
graph isomorphism (GI), the  problem of determining whether two graphs
on the  same vertex set  are isomorphic; here the witness is
a permutation  of the vertices that makes  the two graphs equal.

 A problem  $P$  is in
complexity  class   {\bf  co-NP}  if   and  only  if   its  complement
$\overline{P}$ is  in {\bf  NP}. In simple  terms, {\bf co-NP}  is the
class  of   problems  for  which  efficiently   verifiable  proofs  of
``no"-instances, also  called {\em counterexamples}, exist.  
The   complement  of   GI,  graph
non-isomorphism, is clearly an example of a problem in {\bf co-NP}.
GI is not known  to   be  in   {\bf  co-NP}.   

In simple terms, the class of {\bf NP}-complete 
problems is the class of hardest problems in
{\bf NP}  in the  sense that if  one can  find a way  to solve  a {\bf
NP}-complete problem  `easily' (i.e., in polynomial time), then  one can use
that algorithm to solve all  {\bf NP} problems easily \cite{npc}.  The
Boolean satisfiability problem (SAT)  is {\bf NP}-complete
(Given  a boolean  expression, is  there  at least  one assignment  of
true/false values  to the variables  that makes the expression  have a
``true''  value?).  Closely  related to  the SAT  problem is  the TAUT
problem, the problem of determining whether a given boolean formula is
a  tautology,   which  is   {\bf  co-NP}-complete  (Given   a  boolean
expression, does every possible assignment of true/false values to the
variables yields a true statement?).   {\bf BQP} is suspected, but not
known, to be disjoint from the class of {\bf NP}-complete problems and
a strict superset of {\bf P}.  Integer factorization and discrete log,
which are in {\bf BQP}, are  {\bf NP} problems suspected to be outside
of  both  the  class {\bf  P}  and  and  also  of  the class  of  {\bf
NP}-complete problems.   Clearly, ${\bf  P} \subseteq {\bf  NP}$ since
easy  solvability implies easy  verifiability. Intuitively,  one would
expect that the converse is not true.  Yet, interestingly, this has in
fact  not been  proved so  far.  This  encapsulates the  \mbox{{\bf P}
$\stackrel{?}{=} {\bf NP}$} problem,  the most famous open question in
computer science.

Computational  complexity  class \#{\bf  P}  is  the  set of  counting
problems associated  with the decision  problems in {\bf  NP}. That
is, it is the class of function problems of the form ``compute $f(x)$",
where $f$ is the number of accepting paths of an {\bf NP} machine.  Unlike
most  well-known complexity  classes, it  is not  a class  of decision
problems,  but a  class of  function problems.  The canonical \#{\bf  P} 
problem is \#SAT: given a Boolean formula, compute how many satisfying
assignments it has \cite{cpa}.

{\bf PSPACE}  is the set of  decision problems solvable on  a TM using
polynomial  amount of  memory. It is immediate
that ${\bf P} \subseteq {\bf PSPACE}$,
since a poly-time algorithm can consume only polynomial space.
Given a  boolean  formula using  only
$\exists$  (existential) quantifiers  to bind  the variables  yields a
problem in SAT; using  only $\forall$ (universal) quantifiers yields a
problem in TAUT.  Alternating  both yields a
True Quantified Boolean
formula (TQBF)  problem, which is {\bf  PSPACE}-complete.  
A decision problem is in {\bf PSPACE}-complete if it is in 
{\bf PSPACE}, and every problem in {\bf PSPACE} can be reduced 
to it in polynomial time. For example, a SAT problem
is the question of whether the following is true:
\mbox{$\exists_{x_1}\exists_{x_2}\exists_{x_3}\exists_{x_4}  (x_1 \vee
\neg x_3 \vee  x_4) \wedge (\neg x_2 \vee x_3 \vee  \neg x_4)$}.
This can be generalized      to a TQBF by replacing the above quantified
Boolean formula by
\mbox{$\exists_{x_1}\forall_{x_2}\exists_{x_3}\forall_{x_4}  (x_1 \vee
\neg x_3 \vee x_4) \wedge (\neg x_2 \vee x_3 \vee \neg x_4)$}.
It is known that ${\bf NP} \subseteq {\bf PSPACE}$.
The following containments are known to hold:
${\bf P} \subseteq {\bf BPP} \subseteq {\bf BQP} \subseteq 
{\bf PP} \subseteq {\bf PSPACE}$ and
${\bf P} \subseteq {\bf NP} \subseteq {\bf PP} \subseteq {\bf PSPACE}$
\cite{pp}.

In the remaining part of this Section, we consider variants
of QM that lead to more powerful (in the sense of 
complexity) models of computation.
QM is known to be an `island in theoryspace':
it is strictly linear, unitary and having measurements obey 
the Born $|\psi|^2$ rule \cite{aar04}.
One cannot give up even one of these features, without 
collapsing its whole structure, as viewed from some physical or
computational perspective. 
In this sense, it is unlike, for example, gravity, where one 
can define a family of Brans-Dicke theories in the neighborhood of
General Relativity that are practically indistinguishable from the
latter. 

Consider solving  SAT on a computer  powered by nonlinear  QM.  
This is easily solved if we have a polynomial time algorithm
that determines whether there exists an input value $x$ for which $f(x) = 1$,
where $f$ is a boolean black box function.
To begin  with, we assume $f(x)=1$  on at most
one   value    of   $x$.    Prepare   the    state   $|\psi\rangle   =
2^{-n/2}\sum_{x=0}^{2^n-1}    |x\rangle_{\rm   index}|f(x)\rangle_{\rm
flag}$ on $n$ qubits and a `flag' qubit.

There are $2^{n-1}$ 4-dim  subspaces, consisting of the first index
qubit and the  flag qubit, labelled by the  index qubits $2,\cdots,n$.
On each such subspace, the first index qubit and flag qubit are in one
of the  states $|00\rangle  + |11\rangle$, $|01\rangle  + |10\rangle$,
$|00\rangle + |10\rangle$.   A `nonlinear OR' is applied  to these two
qubits to transform them according to:
\begin{equation}
\label{eq:nonlinor}
\left.
\begin{array}{ll}
|00\rangle + |11\rangle \\
|01\rangle + |10\rangle \\
|01\rangle + |11\rangle \\
\end{array} \right\} \longrightarrow
|01\rangle + |11\rangle; ~~
|00\rangle + |10\rangle 
\longrightarrow |00\rangle + |10\rangle).
\end{equation}
This operation is repeated $(n-1)$ times, pairing each other 
index qubit with the flag. The number
of terms with 1 on the flag bit doubles with each operation so that
after the $n$ operations, it becomes disentangled and can then
be read off to obtain the answer \cite{bram98}.
A slight modification of this algorithm solves problems in $\#{\bf P}$ 
efficiently. One replaces the flag qubit with $\log_2 n$ qubits
and the 1-bit nonlinear OR with the corresponding nonlinear counting.
The final read-out is then the number of solutions to $f(x)=1$.

One can also solve SAT efficiently via non-unitary QM \cite{aar05}.
For example, to the second register of $|\psi\rangle  = 2^{-n/2}\sum_x|x
\rangle|f(x)\rangle$, apply the nonunitary but invertible gate
\begin{equation}
\label{eq:g}
G = \left( \begin{array}{ll} 2^{-2n} & 0 \\ 0 & 1 \end{array}\right)
\end{equation}
Measurement on the 
second register allows one  to know whether there  exists $x$ 
such that $f(x)=1$  with exponentially small uncertainty.
This also solves the complement of SAT, to which TAUT is 
(polynomial-time many-one) reducible.

As another variant of QM,
suppose  QM allows the probability  of measurement  outcomes to  depend on
other norms $p$ than the 2-norm of Born's $|\psi|^2$ rule. Restricting
the dynamics to be norm-preserving leaves only the trivial dynamics of
generalized permutation matrices.   So the only option seems  to be to
use   manual    normalization:   to stipulate
that when   a    state   $|\psi\rangle   =
\sum_x\alpha_x|x\rangle$ is  measured in the  computational basis, the
probability  of outcome  $x$ is  $|\alpha_x|^p/\sum_y|\alpha_y|^p$.  Since
here norm  is not  required to  be preserved,  the dynamics  is free  to be
unitary or simply consist of  invertible matrices. In the latter case,
`local normalization'  can be an  option. Each of these  three options
can be shown to allow quantum computers to
solve even {\bf PP}-complete problems \cite{pp}
in polynomial time \cite{aar05},  
which  are  believed  to  be  harder  than  {\bf NP}-complete problems.

Similarly,  allowing for  nonlocal signalling  also  permits efficient
solving of SAT.  To see this, observe that  nonlocal signalling almost
always implies a departure from standard QM: nonlinearity 
(some instances include those discussed
in Refs.  \cite{gis90,pol91}), non-unitarity, etc.
As an example of the latter case, 
we note that the gate in Eq. (\ref{eq:g}) can be used
to transmit a nonlocal signal. To do so, 
sender  Alice applies either $G$ or $XGX$
to  her  qubit  in  the  entangled  state  $(1/\sqrt{2})(|01\rangle  +
|10\rangle)$,  shared  between her  and  Bob.  Accordingly, Bob  finds
$|0\rangle$  or $|1\rangle$  with probability  exponentially  close to
1. Here $X$  is Pauli $X$ operator.  This  nonlocal signal
`instantaneously' transmits 
classical information without requiring material or energy transfer
using only the Einstein-Podolsky-Rosen channel \cite{epr}.
It is not surprising that nonlocal signalling power is closely related to
the power to solve  hard  problems efficiently,  inasmuch as  a  similar
``communication  across superposition  branches" is  required  in both
cases. 

In  fact nonlinear  quantum computers  can even  solve  ${\bf PSPACE}$
problems  efficiently \cite{aar05}. To  solve the  {\bf PSPACE}-complete
problem  mentioned above, one  alternatively applies  nonlinear OR's
and AND's, (instead  of only nonlinear OR's as used  to solve SAT) to:
$|\psi\rangle            =            \sum_{x=0}^{2^n-1}|x\rangle_{\rm
index}|f(x)\rangle_{\rm flag}$,  starting with $n$th  and flag qubits,
moving the control  bit sequentially  leftward. Here the one-bit 
nonlinear AND is analogous to Eq. (\ref{eq:nonlinor}), given by:
\begin{equation}
\label{eq:nonlinand}
\left.
\begin{array}{ll}
|00\rangle + |10\rangle \\
|00\rangle + |11\rangle \\
|01\rangle + |10\rangle \\
\end{array}\right\} \longrightarrow
|00\rangle + |10\rangle; ~~
|01\rangle + |11\rangle 
\longrightarrow |01\rangle + |11\rangle).
\end{equation}
This will  disentangle the
flagbit which is then read off to obtain the answer.

The above observations raise the question as to why QM `chooses' to be
such an island in theoryspace-- strictly linear, unitary, local
and conforming to the Born $|\psi|^2$ rule.
A similar  observation  applies to  other promising  candidates
among natural  processes that potentially offer  more efficient models
of  computation, such  as  simulated annealing,  soap bubbles,  protein
folding and `relativistic computation' \cite{aar05}.  On closer inspection,
the  evidence in  support  of their ability to efficiently solve
hard problems is not  found to  be
persuasive. Their  route to efficiency  seems always to be  blocked by
such  features   as  taking  longer  relaxation   or  evolution  times
\cite{zni06}, ending up in local minima, requiring exponentially large
energy, etc.

These considerations lend support  to the {\bf NP}-hardness assumption
(NPHA): that  {\bf NP}-complete problems are
intractable in the physical world \cite{aar05}.  Our confidence in the
probable  veracity of this  assumption stems  not only  from empirical
knowledge  of  the  physical   world,  but  from  noting  that  simple
modifications  to the laws  of (quantum) physics,  which could  have 
led  to the possibility of more effecient computing machines, are not 
found to be availed of in nature.

\section{Computatability considerations \label{sec:turing}} 

Related to the issue of computational complexity
is the question of computability, that is, the existence
of an algorithm to solve a given computational task.
An existential proof for uncomputable functions 
is based  on a counting argument:
the number  of functions $f:  {\bf N} \mapsto \{0,1\}$  is uncountably
many  ($2^{\aleph_0}$), whereas the  number of  TMs is  only countably
infinite.  Thus, most functions are (Turing) uncomputable.
A specific example is Turing's halting problem, which is undecidable. 

Suppose all TMs
are  uniquely numbered as  $M_j(\cdot)$ ($j  = 0,1,2\cdots$)  according to
some consistent  scheme.  Consider the  halting set $H \equiv  \{j ~|~
M_j(j){\downarrow}\}$, consisting of machine  numbers of TMs that halt
when they get as input their  own number.   
Simply running  $M_j(j)$ until it
halts constitutes  an algorithm to accept $H$, that is, to determine  
any ``yes'' instance  to the problem of  
whether $j  \in H$.  Thus $H$  is  semi-decidable or
recursively  enumerable (r.e.). The latter name derives from
the fact that there is
an enumeration procedure (employing  a  `dovetailing'   principle) 
whereby every element in $H$ is eventually detected.

But $H$ is  not co-r.e. (i.e., its complement
$\bar{H}$ is not  r.e) because there is
in general no algorithm to check the ``no'' instance of this question.
Intuitively,  this is because  if a  program does  not halt,  we would
never know  that it  won't do  so at a  later time.   More rigorously,
suppose $\bar{H}$ is r.e: let $d$  be the machine or program number of the
TM  that accepts  $\bar{H}$.  Thus  $n \notin  H ~\Longleftrightarrow~
M_d(n){\downarrow}.$  Therefore:  $d  \notin  H  ~\Longleftrightarrow~
M_d(d){\downarrow}$. However  the definition of $H$ tells  us that: $d
\in H  ~\Longleftrightarrow~ M_d(d){\downarrow}$.  Thus  we have that:
$d  \in H  ~\Longleftrightarrow~  d \notin  H$,  a contradiction.   It
follows that $\bar{H}$  is not r.e. $H$ is  thus non-recursive-- there
exists no general membership algorithm for $H$. 
The halting  function $h(x)
\equiv  \{x  |  x   \in  H\}$  is  thus  algorithmically  uncomputable.
Uncomputability  implies \cite{cha04} G\"odel incompleteness
\cite{God}.

One might ask whether uncomputability  is a limitation of the TM model
of computation, and  whether perhaps an algorithm may  be more general
than  a TM.   According to  the {\em Church-Turing  thesis} (CTT),  the
answer is in  the negative.  CTT asserts that any  problem that may be
intuitively considered as computable  
(in  a  reasonable  sense) is  computable  on  a
TM. That is, the formal concept of a TM captures exactly the 
intuitive idea of an 
algorithm or an {\em effective procedure}.
Note  that  it is not provable,  since it relates
an intuition to a formal notion. 
Nevertheless, CTT is falsifiable in the sense
that it can be refuted by the  discovery  of an intuitively acceptable algorithm
or, more starkly, of some  physical process, for a 
Turing-uncomputable task.

QM is characterized by  infinite parallelism  and the infinite  precision of
amplitudes (the continuum  nature of Hilbert space)  \cite{niels}.  By the
counting  argument, the  cardinality of  the set  of  quantum 
TMs (or programs)
equals that of  all functions $f: {\bf N}  \mapsto \{0,1\}$.  Thus,
the counting argument cannot be used to exclude a {\em quantum} 
algorithm from computing the halting function $h(x)$.
In particular, one can  conceive of a  quantum machine
${\cal  Q}$ that accepts $\bar{H}$: 
$n  \notin   H  ~\Longleftrightarrow~  {\cal
Q}(n){\downarrow}$.  Contradiction through self-reference
is  averted  because ${\cal  Q}$,
being represented by a real number, cannot be the argument to any TM.
(Actually, this argument can be applied also to real-valued TMs.)
 
Yet, it is usually believed that quantum Turing machines (QTM) 
can only compute the same functions that are 
computable with classical Turing machines. 
The QTM model, defined by Deutsch \cite{deutsch} and further formalized by 
Bernstein and Vazirani \cite{umesh}, is
simulable   by  classical  Turing  machines (albeit at the expense  of
exponential  slowdown),
therefore so far as computability  is concerned and within the scope of
this QTM model, the set of computable functions remains the same as that for
(classical) TM. Empirical evidence
suggests that in computers
based on the  principle of relaxing to an  energy minimum that encodes
the  solution, as  DNA  computer, soap  bubbles, simulated  annealing,
etc., the physical relaxation time,  which is a measure of 
computational complexity,
tends to diverge as a function of problem size. 
Another possible impediment to super-Turing power is
noise, which can render unfeasible infinitely precise computation.

We remark  on a further  point: that, even with  infinite parallelism,
quantum  computers  may   need  nonlinearity  to  solve  non-recursive
problems.  This is  analogous  to the difficulty of simulating
nondeterminism using the exponential
parallelism of QTMs  \cite{ben97}.   To see
this, we briefly describe a model of infinite quantum parallelism.

We are given a problem $L \subseteq {\bf N}$ and an 
infinite dimensional quantum system, whose Hilbert space ${\cal H}_S$
is spanned by vectors $\{|j\rangle\}$. Suppose we can engineer a unitary 
transformation $U$ whose action on ${\cal H}_S \otimes {\cal H}_R$,
where ${\cal H}_R$ is the 2-dimensional space of an ancilla, is
such that $U|j\rangle|0\rangle = |j\rangle|f(j)\rangle$, where 
$f(j)= 1$ if $j \in L$ and
$f(j)=0$ otherwise. We assume further that we can arrange
energetically so that all superpositional pathways $j$ terminate within
finitely bounded time $T$ independent of $j$ (this requirement is analogous
to invoking the linear speed-up lemma \cite{hop02} in the model of infinite
parallelism proposed in Ref. \cite{zie04}).

Consider trying to solve 
the  halting   problem,  or,  equivalently,   Hilbert's  10th  problem
\cite{mat71}. In the former case, the action of $U$ on pathway $j$ 
may simulate a given program $P$ 
acting on its input $x$, through the first $j$ steps, and evaluate $f(j)$
to ``1" iff $P(x)$ halts within these steps. In
the  latter problem,  the quantum computer computes on each pathway $j$,
the given Diophantine equation
$D$ on input $j$ and evaluates $f(x)$ to ``1" iff $D(j)=0$.

If $U$ were physically realizable, running this quantum computer on a
superposition of all possible inputs, by virtue of quantum mechanical
linearity, one obtains the possibly entangled 
state $U(\sum_{j=0}^{\infty} |j\rangle|0\rangle) \longrightarrow
\sum_{j=0}^{\infty} |j\rangle|f(j)\rangle$.
To be able to compute Turing-uncomputable functions, we require
a finite method to detect an acceptance on at least one term in the
output superposition of $U$. 
Unfortunately, we can do no better than to quantum search
through all the infinite terms.
We thus recover uncomputability, since a quantum search can yield no
better than a quadratic speedup \cite{ben97}, so that a negative  
answer to $L$ can never  be ruled  out in
this way  in finite  time.  This line  of reasoning suggests that
infinite quantum parallelism 
cannot necessarily be harnessed to solve Turing uncomputable problems.

\section{The quantum measurement problem \label{sec:q}}

The conclusions of the preceding two sections demonstrate that CTT and
NPHA present  us with what is  arguably a twofold  coincidence: (a) On
the one hand,  we find that {\bf NP}-complete problems  do not seem to
be  efficiently solvable  by any  physical means;  (b) on
the other hand, algorithmically
uncomputable problems  do not  seem to be  computable by  any physical
means.  

This \underline{c}lose \underline{c}orrespondence between  the 
efficiency and
power  of  theoretical  \underline{a}lgorithms   
and  those of  \underline{p}hysical  computers 
(abbreviated to CCAP) evokes the  question: How  do we
account   it?    Three  broad possibilities present themselves:
\begin{description}
\item{(1)} that the laws of physics support super-Turing efficiency 
and power, 
but instances of violation of NPHA and CTT in Nature remain unidentified.
\item{(2)}  The   universe  is   not  Turing  computable,   let  alone
polynomial,   but   super-Turing    efficiency   or   power   is   not
``harnessable", because  of quantum mechanical  `insularity' (i.e., QM
being linear, unitary, etc.),  or ``accidents" such noise, energetics,
etc. One possibility that realizes this option is that the universe is
a  quantum   computer,  where   the  origin  of   uncomputabality  and
intractability could be the infinite quantum parallelism (eg., as in a
harmonic oscillator) and exponential quantum parallelism (eg., as in a
multi-qubit system), respectively, with the purported impossibility of
solving the  halting problem or  efficiently simulating nondeterminism
having its origin in the special, `insular' structure of QM.
\item{(3)}  The  universe  is  algorithmic: the states  and  evolution  of
physical  objects are manifestations of discrete  information  stored  and
computations   performed   at    the   `sub-physical'   level   by   a
probabilistic TM, which serves as a `meta-universe'.   
Physical  laws   are  manifestations  of  efficient
sub-physical algorithms  on the  probabilistic TM.  Physical  reality is
fundamentally information theoretic, in  the sense of Wheeler's phrase
``it from bit". We need to clarify, which we do below, why this option
does not contradict the fact that {\bf BQP} apparently characterizes
the observed power of quantum computers.
\end{description}

Are there other  observations that can influence our  choice of one or
other option?  We claim that the {\em quantum measurement problem}
(QMP)  is  one such.   We  will argue  that  QMP  clarifies the  above
situation in  two ways:  first, it  will enable us  to argue  that the
apparent {\bf  BQP}ness of microscopic
quantum states that we are sure we can prepare does not
contradict the proposition that classical macrosystems exist
in {\bf BPP}.  This is
crucial to  justify option (3) as  an explanation of
CCAP;  second, it will enable us to
argue that macro-classicality undermines both options (1) and (2).

QMP is a fundamental problem of interpretting QM, that,
in its simplest form, is concerned with the questions (among others):
Why is the macro-world classical? Why do we not find conspicuous macroscopic
manifestations of quantum interference?  

In slightly more detail: when a quantum measurement 
is performed on a system $S$ in the state $|\psi\rangle
= \sum_j \alpha_j|j\rangle$ ($\sum_j |\alpha_j|^2=1$), quantum 
mechanical linearity implies that a Schr\"odinger
cat state should be generated as follows:
\begin{equation}
\label{eq:qmp}
\sum_j \alpha_j|j\rangle|R\rangle
~\longrightarrow~ \sum_j \alpha_j|j\rangle|R_j\rangle,
\end{equation}
where  $|R\rangle$  is  the  `ready' state,  and  $|R_j\rangle$'s  the
correlated  states of the  measuring apparatus.   Instead, selectively
speaking  (i.e.,   conditioned  on  the  read-out   of  the  measuring
apparatus), only  one of the  possible outcomes $|j\rangle|R_j\rangle$
is observed.  The origin of this apparently discontinuous, non-unitary
jump,  sometimes  called  the   ``collapse  of  the  wavefunction"  or
``reduction  of the  state vector",  is  QMP (or,  more precisely,  an
important part of QMP).

QMP   is   a  long-standing   and   contentious   problem  about   the
interpretation of  QM, with many proposed,  interesting solutions (cf.
\cite{max03} for a detailed review). 
Often the formal collapse is treated only as 
an interpretational concept rather
than an objective physical phenomenon. Part of the reason is that it is 
hard to establish a critical size at which the physical collapse 
may be said to happen or
to come up with a clearly testable mechanism. In the view of many,
it is not ruled out that QMP may be resolved without invoking collapse.
Still, in Refs.  \cite{sri03,sri06}, we
argued that  the mechanism of  environmental decoherence 
\cite{decoh1,decoh2,decoh3},
complemented  by  a wavefunction  collapse,  is  a  reasonable way  to
resolve QMP.  Our support for option  (3) over options (1)  and (2) is
based on this line of thought, in particular, the assumption that 
collapse is an objective occurance.   

Under  option   (3),  one  is   required  to  reconcile   the  assumed
polynomiality (i.e,  {\bf BPP}ness) of the universe  with the observed
exponentiality  of  the quantum  states,  which  empowers the  massive
quantum  parallelism behind Shor's  algorithm \cite{shor}.  The latter
gives a  polynomial time quantum algorithm for  the factoring problem,
which is not believed  to be in {\bf P}.  A crucial  fact here is that
{\em this  exponentiality has never been seen  on macro-scales}, which
of course is QMP.  Therefore, if we accept option (3), the only way to
avoid  the contradiction  between  the presumed  polynomiality of  the
universe  and the  apparent exponentiality  confirmed at  small scales
seems to  be to fundamentally upper-bound the  degree of superposition
allowed for  any quantum system.  By choosing  this bound judiciously,
that  is, by  positioning it  at the  Heisenberg cut  (the  scale that
seemingly  separates the quantum  realm from  the classical),  one can
guarantee that sufficiently small systems will manifest superpositions
and hence  exponential behaviour; on the other  hand, for sufficiently
large  systems,   exponential  (superpositional)  behaviour   will  be
replaced by polynomial (classical)  behaviour.  With this proviso, the
fact that quantum computers  are apparently characterized by {\bf BQP}
at microscopic  scales would not be incompatible  with the proposition
that the universe is in {\bf BPP}.

Conversely, this may  be expressed as follows. We treat  QMP 
interpreted as an objective collapse as a {\em
threshold phenomenon},  which separates  the quantum behaviour  in the
micro-world from  the classical  behaviour in the  macro-world.  
It may be thought of as a sort of {\em quantitative} (as against
qualitative) ``Sure/Shor" separator \cite{saar}, a criterion that
separates the quantum states are that are surely experimentally
preparable, from states that arise in a {\em large-scale} implementation
of Shor's algorithm. Given
the  polynomiality of  the universe  per  option (3),  and the  tested
exponentiality  at small scales,  option (3)  {\em predicts}  that the
quantum  superposition  principle should  break  down  at some  finite
scale.    By   identifying   this   threshold scale   with   the
quantum-classical divide, we have a natural resolution to QMP.  From
this  viewpoint,  macro-classicality  is  a  sign  that the universe
is in {\bf BPP}. 

QMP  interpreted as  a  collapse phenomenon  in  fact undermines  both
options (1) and (2). For example, consider the assumption of linearity
of QM  under option (2) as  a possible explanation  for CCAP.  Clearly,
the validity  of the linearity of  quantum evolution at  all scales is
called  into   question  if  lack   of  superpositions  of   the  type
(\ref{eq:qmp}) is due to collapse.  This contradiction is not weakened
by taking into consideration  noise, measuring errors and such details
\cite{bas00}.   (A caveat  is that  alternative interpretations  of QM
like Many-worlds  or Bohmian may  be exempt from  this contradiction.)
If we regard macro-classicality as a sign of breakdown of linearity at
some scale, then option (2) seems to be disfavored.

This breakdown would also imply that a superposition of infinite terms
is  disallowed. In the  ITS model,  infinite parallelism  is necessary
(though  not   sufficient)  for  computing   non-recursive  functions.
According  to  this model,  the  breakdown in  linearity/superposition
principle also disfavors option (1).

Adopting  option (3)  as the  explanation of  CCAP, we  are led  to the
worldview  that the  universe is  computable,  `simulated/computed' by
{\em  efficient} algorithms  run on  a sub-physical  probabilistic TM,
which serves as a {\em meta-universe}.  Quantum randomness is accepted
as fundamental.  Thus the universe  is described as a polynomial place
in  {\bf  BPP}. We  believe  that option  (3)  is  potentially a  more
`natural'  and `deeper'  explanation of  CCAP than  the  other options.
First, we note that NPHA and CTT follow immediately and naturally: CTT
is simply the consequence of Turing-uncomputability; NPHA follows from
the  situation that probably  no efficient  algorithms exist  to solve
hard  problems (if  such exist,  perhaps  they are  so unobvious  that
Nature  hasn't yet  `discovered'  them!)  Under  option  (3), we  then
conclude   that  Schr\"odinger  evolution   is  linear,   unitary  and
compatible with  no-signalling and quantum measurement  obeys the Born
rule  because if  it were  not so,  the resulting  computational power
would be larger than that  supportable by an algorithmic universe.  In
fact, wavefunction collapse implies  a breakdown in both linearity and
unitarity, but not of a kind that would allow efficient computation of
hard problems,  or, for the  matter, nonlocal signaling.  (though {\em
non-selectively}, that  is, at  the level of  the density  matrix, the
evolution is still linear, and can  be regarded as unitary in a larger
Hilbert  space  \cite{nc00}.) This  suggests  that  while the  insular
structure of  QM in theory  space is special,  a departure of  QM from
insularity may be allowed if  it is compatible with the algorithmicity
of the universe.

These observations allow us to reduce a host of physical laws to basic
results in  computation theory. Furthermore,  teleologically speaking,
there  is  an  obvious,  intrinsic  `motivation'  for  an  algorithmic
universe  to employ efficient  algorithms: if  they indeed  suffice to
engender a sufficiently complex universe  (which seems to be the case,
cf.  Ref.  \cite{ulf}), then  no further `computational effort' on the
universe's part is needed!

Options (1) and (2)  are less satisfactory as fundamental explanations
of CCAP.  For example, consider the assumption of linearity of QM under
option  (2) as a  possible explanation.   There is  no {\em  a priori}
reason to expect that a self-consistent QM should be linear.  Invoking
the prohibition on nonlocal signaling to impose linearity brings in an
extraneous physical criterion. Moreover, it means that one of them has
to accepted axiomatically.

From  the   viewpoint  of  classical computation,  option (3)  is 
conservative because it sides with the belief that
it is unlikely that Nature would manipulate or maintain 
exponentially (not to mention, infinitely) 
large objects ``free of cost" (or, at unit cost)
\cite{oded}. It is intuitively satisfying to picture the 
laws of physics as algorithms for physical dynamics, and hence that
the limits on efficient computation and on computability in the
physical world derived from them, to
correspond to formal, purely mathematical notions of computation. 
Option (3) thus arguably supports the  philosophy that
mathematical truths do not depend  on the laws of physics and suggests
that  insights from  computer science  can  be used  to constrain  the
search for new physics.  In contrast, option (2) arguably supports the
philosophy that  the limits of  mathematics are dependent  on physical
laws.  This  viewpoint  does  not  encourage the  hope  that  computer
scientific  insights  may  constrain  physics, but  instead  that  new
physics may extend the limits of algorithms and mathematics.

An  important  objection to  this  argument  is  the following:  that,
despite  its origin  in physics,  {\bf  BQP} is  a fully  mathematical
notion,  since  the  underlying  concepts,  namely  the  superposition
principle and the tensor product structure of Hilbert space, are fully
mathematical; and that, a  classical computer scientist, accustomed to
a   different  mathematical   framework  (namely   that   of  discrete
mathematics with composite systems described in terms of the cartesian
product) may have found quantum computation a little unfamiliar, simply
because of  unfamiliar mathematics,  rather than because  they involve
physics in any essential way.

This point  merits further consideration,  but we briefly
note  the following: that, unless one adopts option (3), 
one probably has no way, except by empirical
observation of the physical world, 
to fix what one would regard as the most powerful possible 
``reasonable" model of computation, both in the sense of computational 
complexity and of computability.
For example, to one who adopts the viewpoint of option
(1) or (2), there would be no
fundamental explanation for his claim that quantum  computers can exist in
nature, but machines such as (say) nondeterministic computers cannot. 

It  remains  to concretize  option  (3)  in the  form  of  a model  of
measurement that accounts for a bounded degree of superposition.  Such
a model had  been proposed by us earlier  \cite{sri03,sri06}, which we
briefly review in the next Section.

\section{Computational model for Quantum Measurement\label{sec:cmqm}}
This section briefly discusses  a model of quantum measurement, called
the  computational model for  quantum measurement  (CMQM), which  is a
particular realization of option (3).  A fundamental assumption of the
CMQM is  that Hilbert space is discrete  \cite{sri03,sri06}.  The idea
of a discrete Hilbert space  has been independently arrived at in Ref.
\cite{bun05}  on quantum gravity  grounds.  Here  discretization means
that  the  description of  any  quantum  state  with respect  to  some
reference basis in a finite  dimensional Hilbert space requires only a
finite  number ($\mu$)  of bits.   We denote  by ${\cal  H}_{\mu}$ the
Hilbert space ${\cal H}$ discretized at $\mu$-bit accuracy.  Parameter
$\mu$ specifies  an intrinsic limit  on the resolution of  states, and
not  an effective  limit  due  to practical  constraints.   In a  more
detailed model, $\mu$  need not be fixed, but  only upper-bounded, and
discretization may not mean a lattice structure.

A state $|\psi\rangle  \in {\cal H}_{\mu}$ is described  by $\mu$ bits
per amplitude  ($\mu/2$ for the  real and imaginary parts).   Thus the
full state is specified by $D\mu$ bits, where $D$ is the Hilbert space
dimension (Actually, $(D-1)\mu$ bits suffice because of normalization.
However,   for  simplicity,   we  will   ignore  this   detail.)   The
sub-physical computational rate  corresponding to a system's evolution
when  driven  by  a  Hamiltonian $E_j|j\rangle\langle  j|$  is  ${\cal
F}(D,E_j)        =       \frac{2^{\mu/2}}{\hbar}\sum_jE_j       \equiv
\frac{2^{\mu/2}D\bar{E}}   {\hbar}$   operations   per  second   (ops)
\cite{sri03,sri06}.   Unitarity and  normalization hold  true  only to
$\mu$-bit precision. In  principle, continuous $SU(N)$ group structure
can be obtained  in the long wavelength limit  from discrete symmetry.
Thus,  the discretization  is  not necessarily  inconsistent with  the
observed apparent continuous  evolution of quantum   systems
\cite{bun05}.

A consequence of finite $\mu$ is that the degree of superposition of a
coherently evolving  system is bounded  above by $D_{\max}  = 2^{\mu}$
since  in  a   larger  Hilbert  space,  not  all   amplitudes  can  be
resolved. Therefore, the coherent evolution of any physical system can
proceed  along  at  most  a  finite  number,  $2^{\mu}$,  of  parallel
superpositional  pathways (terms  in a  coherent  superposition).  The
number  of quantum  TMs or  programs in  CMQM is  thus  only countably
infinite, and  we recover uncomputability.  In  an arbitrary dynamical
situation,  a Hilbert  space of  dimension larger  that  $2^{\mu}$ may
become energetically  available to the  system.  If in  this situation
the  `loss of  probability' through  unresolvability of  amplitudes is
sufficiently small, then the loss is deemed {\em insignificant}.  Eg.,
given  $\alpha$,  for  sufficiently  large  $\mu$,  a  coherent  state
$|\alpha\rangle$ and  its finite equivalent  $|\alpha_{\mu}\rangle$ in
${\cal   H}_{\mu}$  will   be  hardly   distinguishable   in  practice
\cite{sri06}.

On  the  other  hand,   if  via  interactions  large  entanglement  is
generated, then  the unresolvability of  the state, and  the resultant
loss of  amplitude information, are arguably  no longer insignificant.
In Ref. \cite{sri06}, we  introduced a simple entanglement monotone as
a suitable measure of entanglement resolvable at $\mu$ bits ($\mu$-bit
resolvable or  $\mu$-resolvable).  Given a  set $S$ of  particles, let
$S(\rho_j)$ denote  the single particle marginal  entropy to $\mu$-bit
precision and ${\cal  T}$ the set of all  non-vanishing proper subsets
of $S$. We define $\mu$-resolvable entanglement by:
\begin{equation}
\xi^{(N)}_{\mu} = \left\{ 
\begin{array}{ll} \sum_{j=1}^N S(\rho_j)  & {\rm ~if~} (\lambda_+)_y \ge 
2^{-\mu/2} ~\forall~ y \in {\cal T}. \\ 0 & {\rm ~~~~otherwise}, 
\end{array}\right. 
\end{equation}
where  $(\lambda_+)_y$   is  2nd  largest   eigenvalue  (at  $\mu$-bit
precision)    of   ${\rm   tr}_y(|\Psi\rangle\langle\Psi|)    =   {\rm
tr}_{\overline{y}}(|\Psi\rangle\langle\Psi|)$.
The  idea is  that  two  systems are  not
resolvably  entangled if  the  Schmidt representation  of their  joint
state contains only one coefficient resolvable at $\mu$-bit precision.
An $N$-partite  system possesses genuine  $N$-partite $\mu$-resolvable
entanglement only if every bipartite division reveals $\mu$-resolvable
entanglement.  Two systems that are not $\mu$-resolvably entangled are
said to be $\mu$-separable.

Consider a  system of $N$ particles  with $D \ll 2^{\mu}$,  but $D^N >
2^{\mu}$.  When separable, the system's state is resolvable.  But in a
regime of  high interaction, \mbox{$\xi^{(N)}_{\mu} \approx  N\log D >
\mu$},  so  that  the  loss  of  amplitude  information  can  be  {\em
significant}.  Significant unresolvability leads to {\em computational
instability}:  that is,  the sub-physical  simulation of  the physical
system at $\mu$-bit precision  becomes very noisy.  According to CMQM,
`collapse  of the  wavefunction'  is an  error-preventive response  to
computational instability, whereby the  system is abruptly re-set from
massive  entanglement ($\xi  \approx \mu$)  to a  {\em computationally
stable} state (with $\xi \ll \mu$),  which may or may not be a product
state  in terms  of the  most fundamental  degrees of  freedom  of the
system.

Wavefunction  collapse is  thus understood  as an  algorithmic (rather
than dynamic)  process or transition.   It can be shown  that repeated
cycles  of collapse and  episodes of  $\mu$-unitary evolution  lead to
macro-classicality compatible  with the decoherence of  an open system
\cite{sri03,sri06}.   An implication for  quantum computation  is that
asymptotically, the power of QCs is not {\bf BQP} but {\bf BPP}, since
the  degree of superposition  (the degree  of quantum  parallelism) is
upper-bounded  by $2^{\mu}$.  A  quantum computer  of more  than $\mu$
strongly  interacting qubits  will  tend to  collapse rapidly,  hardly
manifesting  non-classical  behaviour.   Ref.  \cite{bun05}  obtain  a
similar result starting from the assumption of discreteness of space.

\section{Relativity and Quantum gravity\label{sec:qg}}
Earlier we noted that the no-signaling feature reduces to
the assumption of an algorithmic universe built on
efficient algorithms. But we observe that no-signaling 
only implies localism
in the sense that any signal should be mediated by material motion.
It does not imply that there is
an upper-bound (namely, $c$) to the speed of material motion, which Relativity
does. It is not clear that these two versions of localism
may be related but we conjecture they are.

Since the  cardinality of space  or time taken  as a continuum  is the
same as that of the set of functions $f: {\bf N} \mapsto \{0,1\}$, the
possibility  seems  to arise  of  analog  computers with  super-Turing
power, at least in a {\em noiseless} classical world.  One can imagine
solving  the  halting  problem  in  this  rather  exotic  fashion:  by
executing the first  step of computation in half  a second, the second
in the next 1/4 second, the third in 1/8 of a second in the subsequent
interval, and so on until at  the end of 1 second, the halting problem
has been  solved!  Also, given the  ability to compute  $x+y, x-y, xy,
x/y$ and  $\lfloor x\rfloor$ in  one step, where  $x$ and $y$  are any
unlimited-precision  real  numbers, {\bf  NP}-complete  and even  {\bf
PSPACE}-complete problems are  classically solvable in polynomial time
\cite{skonhag}.

For this reason, under option (3), we would expect that physical space
and time must also be disrete.  In fact, reasonable grounds lead us to
expect  that  finite $\mu$  implies  discreteness  of  space and  time
\cite{sri06}. Spacetime discreteness is  of course an idea familiar in
certain approaches to quantum gravity \cite{smo3}. As the discreteness
of spacetime rules out space- or time-based analog computers, we again
recover  properties favoring  computability and  polynomiality  of the
universe.

\section{Discussion \label{sec:konklu}}

The present work may be summarized  as an effort to take CCAP seriously
as a  fundamental physical principle. That the  proposition of the
universe's algorithmicity can provide an economic explanation for such
a wide range  of basic physical laws as  quantum mechanical linearity,
unitarity,   signal-locality,   the   Born   rule,   and   macroscopic
classicality  is justification  for  the belief  that  the search  for
fundamental  physical  laws  can  benefit from  examining  constraints
coming from computation theory.   Although many physicists (as against
computer scientists) may  be skeptical, we believe it  is worth taking
this  idea farther  and  asking whether  all  qualitative features  of
physical laws can be reduced  to results in the logical foundations of
mathematics and computation theory.

It  is a deep-rooted  belief of  scientists that  the laws  of physics
should be  unified into a single  deeper law, simply  because it seems
unlikely (though not impossible)  that the universe is fundamentally a
patchwork of independent, basic  laws.  Similarly, we also expect that
the  mathematical  structure  that  physical  laws will  assume  at  a
sufficiently deep level may force  us to resolve our ambiguity towards
such  profound  and  basic  mathematical concepts  as  continuity  and
infinitessimals  in real  analysis,  and those  like  infinity in  the
logical  foundations  of  mathematics  \cite{weyl}.   We  believe  the
present  approach  indicates  one   way  to  address  this  issue,  by
suggesting  a  concrete  connection  between  computation  theory  and
physical  law.   Further,  it   has  important  implications  for  the
philosophy of  mathematics \cite{brussel}.  It is  not unreasonable to
regard   logic,  mathematics  or   classical  computation   theory  as
independent of the  ``accident" of physical laws and  intrinsic to the
``laws of thought",  however one might conceive. them  The discovery of
Shor's celebrated algorithm, when interpretted under option (2), would
seem to  undermine this  belief.  In contrast,  option (3)
tends to affirm it, and gives us confidence to believe that NPHA and
CTT should be true in {\em  any} instance of the universe. It can thus help
constrain the search for new  physical laws on the road to fundamental
theories such as a theory of quantum gravity or string theory.

Are there tests of the the model of the algorithmic universe?
More simply, is it falsifiable? Clearly, any unequivocal proof that 
wavefunction collapse does {\em not} happen will falsify it.
But this may be difficult to test, given that CMQM is hardly distinghishable
from the effect of decoherence nonselectively, and 
experimental tests of decoherence performed thus far
are incapable of differentiating
the effect of decoherence from that of decoherence terminated by
a collapse \cite{sri06}. This is an important issue we hope to address
in the future.

I thank Dr. Piyush Kurur, Prof. J. Gruska, Mr. Sudhir K.
Singh for comments and discussions. I am thankful to the anonymous
referees for constructive suggestions.

\end{document}